

\magnification=\magstep1
\baselineskip=16pt
\nopagenumbers

\line{\hfil IUHET 227}

\bigskip\bigskip
\bigskip\bigskip

\centerline{\bf
ATOMIC SUPERSYMMETRY, OSCILLATORS,}
\medskip
\centerline{\bf
AND THE PENNING TRAP}
\bigskip
\bigskip
\centerline{V. Alan Kosteleck\'y}

\bigskip

\it

{\parindent=0pt\obeylines\everypar={\hfil}
Physics Department
Indiana University
Bloomington, IN  47405, U.S.A.
}
\bigskip
\bigskip
\bigskip

\rm
\noindent

This paper begins with some background information
and a summary of results in atomic supersymmetry.
The connection between the supersymmetric Coulomb and oscillator
problems in arbitrary dimensions is outlined.
Next, I treat the issue of finding a description of
supersymmetry-based quantum-defect theory
in terms of oscillators.
A model with an anharmonic term
that yields analytical eigenfunctions
is introduced to solve this problem in arbitrary dimensions.
Finally,
I show that geonium atoms
(particles contained in a Penning trap)
offer a realization of
a multidimensional harmonic oscillator in an idealized limit.
The anharmonic theory presented here provides a means of modeling
the realistic case.

\bigskip
\bigskip
\bigskip
\centerline{\it
Written in honor of Franco Iachello on his 50th birthday
}
\centerline{\it
Published in the Festschrift volume
}
\centerline{\it
`Dynamic Symmetries and Spectrum-Generating Algebras
in Physics'
}
\centerline{\it
(Symmetries in Science VII, Plenum, New York, 1993)
}

\vfill\eject

\hsize=6.125truein
\vsize=10truein
\voffset=-0.5truein
\baselineskip=12pt
\nopagenumbers
 at 10truept

\def\th{\theta}

\def\rh{\rho}

\def\om{\omega}

\def\fr#1#2{{{#1} \over {#2}}}

\def\lsim{\mathrel{\rlap{\lower4pt\hbox{\hskip1pt$\sim$}}
    \raise1pt\hbox{$<$}}}
\def\gsim{\mathrel{\rlap{\lower4pt\hbox{\hskip1pt$\sim$}}
    \raise1pt\hbox{$>$}}}

\def\Xbc(#1){X^\mu(#1),b(#1),c(#1)}

\def\half{{\textstyle{1\over 2}}}
\def\frac#1#2{{\textstyle{{#1}\over {#2}}}}

\phantom{.}
\vskip 1.4truein

\leftline{\bf ATOMIC SUPERSYMMETRY, OSCILLATORS,}
\leftline{\bf AND THE PENNING TRAP}
\bigskip
\bigskip

{\parindent=68pt

V. Alan Kosteleck\'y

\bigskip

Physics Department

Indiana University

Bloomington, IN 47405 U.S.A.}

\bigskip\bigskip
\leftline{\bf INTRODUCTION}
\bigskip

It is a pleasure to
participate in celebrating Franco Iachello's 50th birthday.
I wish him many happy returns.

At least two physical supersymmetries are known to exist
in nature.
One, discovered by Franco and his coworkers,
makes connections between different nuclei
[1].
The other, atomic supersymmetry,
is the subject of part of this paper.
It can be viewed as a symmetry-based approach
to the construction of an effective central-potential model
describing the behavior of the valence electron in atoms and ions.

This paper begins with some background information
and a summary of results in atomic supersymmetry.
The connection between the supersymmetric Coulomb and oscillator
problems in arbitrary dimensions is outlined.
Next, I treat the issue of finding a description of
supersymmetry-based quantum-defect theory
in terms of oscillators.
A model with an anharmonic term
that yields analytical eigenfunctions
is introduced to solve this problem in arbitrary dimensions.
Finally,
I show that geonium atoms
(particles contained in a Penning trap)
offer a realization of
a multidimensional harmonic oscillator in an idealized limit.
The anharmonic theory presented here provides a means of modeling
the realistic case.

\bigskip\bigskip
\leftline{\bf SUPERSYMMETRIC QUANTUM MECHANICS}
\bigskip

This section provides some background
in supersymmetric quantum mechanics [2] and establishes notation.

A quantum-mechanical hamiltonian $H_S$
is said to be supersymmetric
if it commutes with
$N$ supersymmetry operators $Q_j$
and if it is generated by anticommutators
according to
$$\{Q_j,Q_k\} = \delta_{jk} H_S
\quad .\eqno(1)$$
The operators $H_S$ and $Q_j$
form the generators of a superalgebra
denoted by sqm($N$).

For the purposes of this paper it suffices to consider the
special case $N=2$, with superalgebra sqm(2).
Define the linear combinations
$$Q=\sqrt{\half}~(Q_1+iQ_2)~~,~~~~
Q^\dagger =\sqrt{\half}~(Q_1-iQ_2)
\quad .\eqno(2)$$
Then, the supersymmetric hamiltonian can be written
$$H_S=\{Q,Q^{\dag} \}
\quad .\eqno(3)$$

For one-dimensional quantum systems,
the superalgebra sqm(2) admits a two-dimensional representation.
Write
$$Q=\pmatrix{0&0\cr A&0\cr} ~~,~~~~
H_S= \pmatrix{h_+&0\cr 0&h_-\cr}
\quad ,\eqno(4)$$
with
$$A = -i\partial_x -i U^\prime/2
\quad .\eqno(5)$$
Here,
$U^\prime$ denotes $dU/dx$ for some function $U=U(x)$.
In this representation,
the supersymmetric hamiltonian $H_S$ contains two components
$h_+$ and $h_-$,
referred to as
the bosonic and fermionic hamiltonians, respectively.
These satisfy the equations
$$h_\pm \Psi_{\pm n} \equiv \bigl[ -{{d^2} \over {dx^2}} +
 V_\pm (x) \bigr] \Psi_{\pm n} = \epsilon_n \Psi_{\pm n}
 \, , \eqno(6)$$
with
$$V_\pm (x) =  (\half U^\prime)^2 \mp \half U^{\prime\prime}
\quad .\eqno(7)$$

Using the above relations,
some general properties of
the supersymmetric system can be obtained.
First, for unbroken supersymmetry
the ground-state energy is zero.
Second,
except for the ground state
(which appears in the spectrum of $h_+$)
the bosonic and fermionic spectra are degenerate.
Finally,
the supersymmetry generators $Q, Q^\dagger$
map degenerate states from the two sectors into each other.

\bigskip\bigskip
\leftline{\bf
ATOMIC SUPERSYMMETRY: EXACT LIMIT}
\bigskip

Consider the Schr\"odinger equation for the hydrogen atom.
In spherical polar coordinates,
the equation separates into angular and radial parts.
The angular part gives the spherical harmonics,
while the radial part can be expressed as
$$\bigl[-{{d^2}\over{dy^2}}
- {1\over y} + {{l(l+1)}\over{y^2}}
- {1\over2} E_n
\bigr] \chi_{nl} (y) = 0
\quad .\eqno(8)$$
Here,
atomic units are used, and
$$y=2r~~,~~~~
E_n = -\fr 1 {2n^2}~~,~~~~
\chi_{nl} (2r)=rR_{nl} (r)
\quad .\eqno(9)$$
The radial wave functions $R_{nl}(r)$
are given by
$$R_{nl}(r) = {2\over{n^2}}
 \Bigl[ {{\Gamma(n-l)} \over{\Gamma(n+l+1)}} \Bigr]^{1\over2}
 \Bigl({{2r}\over n} \Bigr)^l \exp \Bigl( -{r\over n} \Bigr)
 L_{n-l-1}^{(2l+1)} \Bigl( {{2r}\over n} \Bigr)
\quad .\eqno(10)$$
In this expression,
$L_n^{(\alpha)} (x)$
are the Sonine-Laguerre polynomials,
rather than the more restricted Laguerre polynomials
(for which $\alpha$ must be integer).
This distinction is important in subsequent sections.
The Sonine-Laguerre polynomials are defined by
$$L_n^{(\alpha)} (x) = \sum_{p=0}^n (-x)^p
 {{\Gamma(n+\alpha+1)}\over{p! \, (n-p)! \, \Gamma(p+\alpha+1)}}
\quad .\eqno(11)$$

With $l$ fixed,
the terms in brackets in Eq.\ (8)
can be reinterpreted in terms of the hamiltonian $h_+$
of Eq.\ (6) as $h_+ - \epsilon_n$.
The requirement that the ground-state energy be zero
permits the separation of $h_+$ and $\epsilon_n$,
so that the supersymmetric partner hamiltonian $h_-$ and the
supersymmetry generator $Q$ can be found
[3].
These can be given by specifying the function $U$ of Eq.\ (5),
which is
$$U(y) = {{y}\over{l+1}} - 2(l+1) \ln y
\quad .\eqno(12)$$
Explicitly,
the hamiltonian $h_-$ looks like $h_+$
but with the constant $l(l+1)$ replaced with $(l+1)(l+2)$,
i.e.,
$$h_- - h_+ = \fr {(2l+1)} {y^2}
\quad .\eqno(13)$$
This shift implies that
for each $l$ the eigenfunctions of $h_-$
are $R_{n,l+1}$, where $n \geq 2$.
Together with the continuum states,
they form a complete and orthonormal set.

This formalism can be given a useful physical interpretation
as follows.
Consider the case $l=0$.
The bosonic sector then describes the s orbitals of hydrogen.
Since the spectrum of $h_-$ is degenerate with that of $h_+$
except for the ground state,
and since the supersymmetry generator $Q$ acts on the radial
part of the hydrogen wavefunctions but leaves the spherical
harmonics untouched,
$h_-$ describes a physical system that appears hydrogenic
but that has the 1s orbitals inaccessible.
One way of realizing this in practice is to fill the 1s orbitals
with electrons,
thereby excluding the valence electron by the Pauli principle.
The element with filled 1s orbitals
and one valence electron is lithium.
This suggests that
$h_-$ should be interpreted as an effective one-body hamiltonian
describing the valence electron of lithium
when it occupies the s orbitals.
At this level,
the description cannot be exact
because most of the electron-electron
interactions are disregarded.
However, these can be introduced as
supersymmetry-breaking terms.
One procedure for this is outlined
in a later section of this paper.
Even in the absence of such terms,
some experimental support for this atomic supersymmetry
can be adduced; see ref. [3].

By redefining the energy of the 2s orbital
in lithium to be zero,
the hamiltonian $h_-$ becomes a suitable choice for a bosonic
hamiltonian of a second supersymmetric quantum mechanics.
The fermionic partner can be constructed,
and an analogous interpretation to the one above can be made.
This suggests
the s orbitals of lithium and sodium should also be viewed as
supersymmetric partners.
The process can be repeated for s orbitals and can also be
applied for other values of the angular quantum number $l$,
leading to supersymmetric connections
among atoms and ions across the periodic table.
In the exact-symmetry limit,
these connections all involve integer shifts in $l$
and are linked to the Pauli principle.
See ref. [3] for more details.

\vfill\eject
\leftline{\bf
OSCILLATOR REFORMULATION: EXACT CASE
}
\bigskip

Before describing
a method for incorporating supersymmetry-breaking effects,
it is appropriate to discuss an alternative formulation of
the exact-symmetry case using harmonic oscillators.
This section outlines the connections that exist between
the radial equations for atomic supersymmetry
generalized to arbitrary dimensions
and those for the supersymmetric harmonic oscillator
[4].

Consider first the $d$-dimensional Coulomb problem.
Upon separation into an angular and a radial part,
the radial equation appears:
$$\Bigl[-{{d^2}\over{dy^2}}
- {1\over y} + {{(l+\gamma)(l+\gamma+1)}\over{y^2}}
- {1\over2} E_{dn}
\Bigr] v_{dnl}(y) = 0
\quad .\eqno(14)$$
Here,
atomic units are used,
and
$$y=2r~~,~~~~
E_{dn} = - \fr 1 {2(n+\gamma)^2}~~,~~~~
\gamma = \half (d-3)
\quad .\eqno(15)$$
The radial wave functions are given by
$$v_{dnl}(y) =
c_{dnl}~ y^{l+\gamma +1} \exp \bigl( -y/2(n+\gamma) \bigr)
L_{n-l-1}^{(2l+2\gamma +1)} \bigl( {y/(n+\gamma)} \bigr)
\quad ,\eqno(16)$$
where $c_{dnl}$ is a normalization constant.

As before,
this one-variable equation can play the role of the bosonic
hamiltonaian in a supersymmetric quantum mechanics.
Appropriately redefining the energy zero so that the ground
state has vanishing eigenvalue permits the identification
of $h_+$ and hence of $U$, $Q$ and $h_-$.
For example,
$$U(y) = {{y}\over{l+\gamma +1}} - 2(l+\gamma +1) \ln y
\quad .\eqno(17)$$
Just as for the $d=3$ case,
the fermionic and bosonic hamiltonians differ only by the
replacement of $l$ by $l+1$,
so that
$$h_- - h_+ = \fr {2(l+\gamma +1)} {y^2}
\quad .\eqno(18)$$
As required,
all the results of the previous section are recovered
if $d=3$, i.e., $\gamma = 0$.

Consider next the supersymmetric harmonic oscillator.
For convenience,
analogous variables to the Coulomb-problem quantities
$y$, $d$, $n$, $l$, $h_\pm$
are now denoted by upper case symbols
$Y$, $D$, $N$, $L$, $H_\pm$.
Thus,
upon separation of the angular and radial parts of the
Schr\"odinger equation for
the $D$-dimensional harmonic oscillator,
the radial equation is obtained as:
$$\Bigl[-{{d^2}\over{dY^2}}
+ Y^2 + {{(L+\Gamma)(L+\Gamma+1)}\over{Y^2}}
- 2E_{DN}
\Bigr] V_{DNL}(Y) = 0
\quad .\eqno(19)$$
Here, atomic units have again been used for simplicity,
and the oscillator is assumed to have unit frequency.
The radial variable is now $Y$,
and
$$E_{DN} = \half (2N+2\Gamma +3)~~,~~~~
\Gamma = (D-3)/2
\quad .\eqno(20)$$
The radial wave functions are given by
$$V_{DNL}(Y) = C_{DNL}~ Y^{L+\Gamma +1} \exp (-Y^2/2)
L_{N/2-L/2}^{(L+\Gamma + 1/2)} (Y^2)
\quad ,\eqno(21)$$
with $C_{DNL}$ a normalization constant.
Note that
the usual expressions for
the harmonic oscillator in three dimensions
are recovered when $\Gamma = 0$.

If Eq.\ (19) is used to define the hamiltonian $H_+$
of a supersymmetric quantum mechanics
(a redefinition of the energy zero is again needed),
then the supersymmetry is specified
by a function $U$ given by
$$U(Y) = Y^2 - 2(L+\Gamma +1) \ln Y
\quad .\eqno(22)$$
It then follows that $H_-$ differs from $H_+$
by the replacement of $L$ with $L+1$.
Therefore,
$$H_- - H_+ = \fr {2(L+\Gamma +1)} {Y^2}
\quad .\eqno(23)$$

So far,
four eigenspectra associated with
the supersymmetric Coulomb and oscillator problems
have been introduced,
defined by the four hamiltonians
$h_+$, $h_-$, $H_+$, and $H_-$.
The hamiltonians
$h_+$ and $h_-$
are related by the map $l \rightarrow l+1$,
and the hamiltonians
$H_+$ and $H_-$
are related by $L \rightarrow L+1$.
The next step is to relate
the $d$-dimensional Coulomb problem
to the $D$-dimensional oscillator,
i.e.,
connect $h_+$ to $H_+$.

It can be shown
[4] that an eigenfunction of $h_+$
can be transformed by a one-parameter mapping
into an eigenfunction of $H_+$.
Explicitly,
the functions $v_{dnl}$ and $V_{DNL}$ are connected by
the equation
$$v_{dnl}\bigl ( (n+\gamma)Y^2 \bigr )
= K_{DNL} Y^{1/2} V_{DNL}(Y)
\quad ,\eqno(24)$$
where $K_{DNL}$ is a proportionality constant and
$$
D=2d-2-2\lambda~~,~~~~
N=2n-2+\lambda~~,~~~~
L=2l+\lambda
\quad .\eqno(25)$$
The integer $\lambda$ is the mapping parameter.
Notice in particular that only
oscillators in even dimensions appear.
Note also that if it is desired
that more than one eigenfunction of $h_+$
be mapped into the eigenspace of a specified $H_+$,
then the possible choices of $D$, $N$, $L$,
and $\lambda$ can become constrained.

The existence of the one-parameter map
between $h_+$ and $H_+$
combined with the supersymmetry maps
evidently establishes connections between
any two of the four hamiltonians
$h_+$, $h_-$, $H_+$, and $H_-$.
More details can be found in ref. [4].

\bigskip\bigskip
\leftline{\bf
BROKEN SUPERSYMMETRY AND QUANTUM-DEFECT THEORY
}
\bigskip

As noted above,
atomic supersymmetry in the exact limit
is not physically realized
because the valence electron interacts
with the core electrons
by more than the Pauli principle.
This section discusses the incorporation
of supersymmetry-breaking
effects in the context of alkali-metal atoms.

One important effect of the interactions
between the valence electron
and the core is the change in energy eigenvalues relative
to the hydrogenic case.
In alkali-metal atoms,
the Rydberg series
[5] provides a simple formula for the measured energies,
given by
$$E_{n^{\ast}} = - \fr 1 {2{n^{\ast}}^2}
\quad .\eqno(26)$$
In this expression,
$$n^{\ast} = n - \delta (n,l)
\quad ,\eqno(27)$$
where $\delta (n,l)$ is called the quantum defect.
For a fixed value of $l$ and increasing $n$,
it turns out that the quantum defects
rapidly attain asymptotic values:
$\delta (n,l) \simeq \delta (l)$.

The changes in the energy eigenvalues imply that the exact
atomic supersymmetry is broken.
The breaking can be viewed
as an additional contribution $H_B$
to the supersymmetric hamiltonian $H_S$ of Eq.\ (4).
For example,
if $h_+$ arises from the radial equation for hydrogen
and $h_-$ is interpreted
as the radial equation for the valence electron
of lithium in the exact-supersymmetry limit,
then the hamiltonian $H$ describing the two systems
\it including \rm supersymmetry-breaking effects
can be taken as
$$H = H_S + H_B
\quad ,\eqno(28)$$
where $H_B$ has the form
$$H_B = \pmatrix{0&0\cr 0&V_B(y)\cr}
\quad\eqno(29)$$
and $V_B(y)$ is such as to generate
the observed energy eigenspectrum of lithium.

The determination of a suitable $V_B$ is not straightforward.
However,
it turns out that a functional form for $V_B$ can be found
that yields analytical eigenfunctions as solutions to the
Schr\"odinger equation
[6].
It is
$$V_B(y) = \fr {l^*(l^*+1) - l(l+1)}{y^2}
+ \fr {n^2 - {n^\ast}^2} {4 n^2 {n^{\ast}}^2}
\quad .\eqno(30)$$
Here,
$l^\ast$ is a modified angular quantum number given by
$$l^{\ast} = l +i(l) - \delta(l)
\quad ,\eqno(31)$$
where $i(l)$ is an integer parameter
shifting the angular quantum number in a manner
characteristic of supersymmetry.
(If desired,
$\delta(l)$ could be replaced by $\delta (n,l)$.)
This model effectively replaces the hydrogenic radial equation
with one of similar form but involving
$n^\ast$ and $l^\ast$ rather than $n$ and $l$.

By construction, the energy eigenvalues are those
of the physical atom.
The resulting eigenfunctions
$R^*_{n^\ast l^\ast}(r)$ are analytical
and are given by
$$R^*_{n^\ast l^\ast}(r) = {2\over{{n^\ast}^2}}
 \Bigl[ {{\Gamma(n^\ast -l^\ast)}
 \over{\Gamma(n^\ast +l^\ast +1)}}
\Bigr]^{1\over2}
\Bigl({{2r}\over n^\ast} \Bigr)^{l^\ast}
\exp \Bigl( -{r\over n^\ast} \Bigr)
 L_{n-l-i-1}^{(2l^\ast+1)} \Bigl( {{2r}\over n^\ast} \Bigr)
\quad .\eqno(32)$$
The Sonine-Laguerre polynomials enter again
because
$${n^{\ast}}-{l^{\ast}}-1 \, = n - l - i(l) -1
\quad\eqno(33)$$
remains integer.
For asymptotic quantum defects
$\delta (l)$ and including the continuum states,
these eigenfunctions form an orthonormal and complete set.

More details about this construction can be found in ref.
[6].
(A connection to parastatistics
is elucidated in ref.
[7].)
There is a reasonable body of evidence to support the notion
that the analytical eigenfunctions
provide a good model for the
valence electron,
especially in alkali-metal atoms.
For instance,
transition probabilities calculated
with the analytical eigenfunctions agree with experiment
and with accepted values
[8].
(Some recursion formulae for matrix elements are given in ref.
[9].)
Transition probabilities
for other elements, notably alkaline-earth ions,
have also been obtained in this way
[10].
Moreover,
these analytical eigenfunctions have been used as
trial wavefunctions in detailed atomic calculations
[11].
Stark maps for the alkali-metal atoms
can also be calculated using
the model
[12].
The resulting clear anticrossings
and small-field quadratic Stark effects
for the s and p orbitals are
in agreement with experiment.
For example,
the model yields Stark maps
for the $n=15$ lines of lithium and sodium
that are indistinguishable from
the numerical and experimental results of ref.
[13].
(Ref.
[12] also studied other possible
quantum-mechanical supersymmetries involving hydrogen.
In particular, a double sqm(2) appears when the separation
is carried out in parabolic coordinates.)
The model is expected to break down at short distances
from the nucleus (of order of the core size),
but despite this
some dominant features
of the fine structure in alkali-metal atoms
are correctly reproduced and
the Land\'e semiempirical formula naturally appears
[14].

\bigskip\bigskip
\leftline{\bf
OSCILLATOR REFORMULATION: BROKEN CASE
}
\bigskip

This section presents a reformulation of
the analytical quantum-defect model
in terms of oscillators with radial equation modified
by an anharmonic term.
For generality,
the connection between the
two is treated in arbitrary dimensions.
The oscillator models that appear
have analytical solutions.
The link between the two theories
is via a three-parameter map.
In the limit of vanishing quantum defect,
this map provides a generalization
of the ones of ref. [4] discussed above.
For example,
it can be used to connect a Coulomb problem
to an anharmonic oscillator with \it odd \rm dimensionality.

The first step is to construct the $d$-dimensional extension
of the quantum-defect model of ref.
[6].
This is done by adding
to the hamiltonian (14) an extra term $V_B^d(y)$
generalizing $V_B$ in Eq.\ (30),
with
$$V_B^d(y) =
\fr {(l^*+\gamma)(l^*+\gamma+1) - (l+\gamma)(l+\gamma+1)} {y^2}
+ \fr {(n+\gamma)^2
- ({n^\ast}+\gamma)^2} {4 (n+\gamma)^2 (n^{\ast}+\gamma)^2}
\quad .\eqno(34)$$
Here,
$n^\ast$ and $l^\ast$ are modified quantum numbers given by
$$n^\ast = n - \delta (d,l)~~,~~~~
l^{\ast} = l +i(d,l) - \delta (d,l)
\quad ,\eqno(35)$$
in analogy with Eqs.\ (27) and (31),
with
$$\gamma = \half (d - 3)
\quad\eqno(36)$$
as before.
The ensuing radial equation in atomic units is
$$\Bigl[-{{d^2}\over{dy^2}}
- {1\over y} + {{(l^\ast+\gamma)(l^\ast+\gamma+1)}\over{y^2}}
- {1\over2} E_{d n^\ast}
\Bigr] v^\ast_{d n^\ast l^\ast}(y) = 0
\quad ,\eqno(37)$$
where
$$E_{d n^\ast} = - \fr 1 {2(n^\ast +\gamma)^2}
\quad .\eqno(38)$$
The radial wave functions solving Eq.\ (37) are given by
$$v^\ast_{d n^\ast l^\ast }(y) =
c^\ast_{d n^\ast l^\ast }~ y^{l^\ast +\gamma  +1}
\exp \bigl( -y/2(n^\ast +\gamma ) \bigr)
L_{n-l-i-1}^{(2l^\ast +2\gamma  +1)}
\bigl( {y/(n^\ast +\gamma )} \bigr)
\quad ,\eqno(39)$$
where $c^\ast_{d n^\ast l^\ast }$ is a normalization constant.

To identify a mapping between the quantum-defect theory
and an oscillator-type model,
a (supersymmetry-breaking) term $V_B^D(Y)$ modifying
the harmonic-oscillator radial hamiltonian (19)
is needed.
A suitable choice is
$$V_B^D(Y) =
\fr {(L^*+\Gamma)(L^*+\Gamma+1) - (L+\Gamma)(L+\Gamma+1)} {Y^2}
+2(N-N^\ast)
\quad .\eqno(40)$$
In this expression,
$$\Gamma = \half (D - 3)
\quad\eqno(41)$$
as before,
and the modified quantum numbers $N^\ast$ and $L^\ast$
are given by
$$N^\ast = N - 2\Delta (D,N,L) ~~,~~~~
L^{\ast} = L +2I(D,L) - 2\Delta(D,N,L)
\quad ,\eqno(42)$$
where the integer $2I(D,L)$ is a supersymmetry-type shift
and $2\Delta$ represents a quantum anharmonicity
(which can be viewed as an oscillator `defect').
The factors of two are introduced
for notational simplicity in what follows.
The extra term (40) introduces an anharmonic piece into the
oscillator potential,
which in turn changes the energy eigenspectrum.
In atomic units with a unit-frequency oscillator,
the anharmonic radial equation becomes
$$\Bigl[-{{d^2}\over{dY^2}}
+ Y^2 + {{(L^\ast+\Gamma)(L^\ast+\Gamma+1)}\over{Y^2}}
- 2E_{D N^\ast}
\Bigr] V^\ast_{D N^\ast L^\ast}(Y) = 0
\quad ,\eqno(43)$$
where the energy eigenvalues are shifted according to
$$E_{D N^\ast} = \half (2N^\ast +2\Gamma +3)
\quad .\eqno(44)$$
The eigensolutions for this anharmonic oscillator are
$$V^\ast_{D N^\ast L^\ast }(Y) =
C^\ast_{D N^\ast L^\ast }~ Y^{L^\ast +\Gamma  +1}
\exp \bigl( -Y^2/2 \bigr)
L_{N/2-L/2-I}^{(L^\ast +\Gamma  +1/2)}
\bigl( {Y^2 } \bigr)
\quad ,\eqno(45)$$
where $C^\ast_{D N^\ast L^\ast}$ is a normalization constant.

The map between the radial equations for
the $d$-dimensional quantum-defect theory
and the $D$-dimensional anharmonic oscillator
connects the eigensolutions $v^\ast_{d n^\ast l^\ast}$
and $V^\ast_{D N^\ast L^\ast}$.
It is given by
$$v^\ast_{d n^\ast l^\ast}\bigl ( (n^\ast+\gamma)Y^2 \bigr )
= K^\ast_{D N^\ast L^\ast} Y^{1/2} V^\ast_{D N^\ast L^\ast}(Y)
\quad ,\eqno(46)$$
where $K^\ast_{D N^\ast L^\ast}$ is
a proportionality constant and
$$
D=2d-2-2\lambda~~,~~~~
N=2n+2(\Delta - \delta) -2+\lambda~~,~~~~
L=2l+2(\Delta - \delta)-2(I-i)+\lambda
\quad .\eqno(47)$$
For fixed $d$, $n$, $l$, $\delta$, and $i$
there are three quantities that
effectively act as mapping parameters:
$\lambda$, $\Delta$, and $I$.
For the eigenfunctions (45) to exist $I$ must be integer,
so the supersymmetry-type shift $2I$ in $L$
is an even integer.
Since $D$ must also be an integer,
$\lambda$ is integer or half-integer.
The quantum numbers $N$ and $L$ are also integer,
which implies that
$2(\Delta - \delta)+\lambda$ must be integer.
Note that this generalizes the exact-symmetry case:
the requirement that $\lambda$ be a whole
integer is no longer needed
because half-integral values can be absorbed in the difference
$2(\Delta - \delta)$.
This means, for example,
that when the Coulomb problem is treated in the exact limit
($\delta = 0$)
it is now possible to map it
into a (modified) oscillator in an \it odd \rm
number of dimensions,
provided $\Delta$ is quarter-integer valued.
As in the exact case,
for maps between \it sets \rm of states
with specified $d$ and $D$,
further restrictions on the possible values
of $D$, $N$, $L$, $I$, $\Delta$
and $\lambda$ may appear.

\bigskip\bigskip
\leftline{\bf
SUPERSYMMETRIC OSCILLATORS AND THE PENNING TRAP
}
\bigskip

In this section,
I demonstrate that geonium atoms provide
a physical realization of
a $D>1$ supersymmetric harmonic oscillator.
For simplicity,
the specific case $D=2$ is considered,
although under suitable conditions
higher values of $D$ may appear.
In practical situations the supersymmetry is broken
for reasons to be described.
The analytical anharmonic oscillator model
introduced in the previous section should provide
a good approximation to the exact wavefunctions
for this case.
Space limitations prevent more than a sketch of the
relevant physics being given here;
details will appear elsewhere.

Geonium atoms are formed by a set of charged particles
bound in a Penning trap [15],
which is a suitable combination of
a homogeneous magnetic field
and an electrostatic quadrupole potential.
The simplest geonium atom
has just one trapped particle
of charge $e$ and mass $m$ [16].
Successively adding further electrons in the trap generates
elements of the geonium periodic table.

For simplicity,
consider the idealized Penning trap
with electromagnetic fields specified
in cylindrical coordinates $(\rh, \th, z)$ by
$${\bf B} = B {\hat z}~~,~~~~\phi =
\half \fr V {d^2} (z^2 - \half \rh^2)
\quad .\eqno(48)$$
The quantity $d$ is a measure of the trap dimension
and is to be specified in terms of the
configuration of the quadrupole electrodes.
For (stable) trapping, $eV > 0$.
The quantum-mechanical motion
of the particle in the field $\bf B$
is that of a harmonic oscillator with binding frequency
equal to the cyclotron frequency,
given in SI units by
$$\omega_c = \fr {\vert e B \vert} {m}
\quad .\eqno(49)$$
(In fact, there are \it two \rm
oscillators involved in this motion,
but only one enters the quantum hamiltonian.)
Similarly,
the electrostatic field generates an axial harmonic motion
independent of the cyclotron motion,
with axial frequency
$$\omega_z = \sqrt{\fr {e V} {m d^2}}
\quad .\eqno(50)$$
These are the motions of primary interest here.

The simultaneous presence of electric and magnetic fields
also generates another (unbound) circular motion,
called the magnetron motion,
with frequency $\om_m$.
For simplicity, this is largely disregarded here.
The eigenvalue spectrum of the system is split by all these
interactions and also (for particles with spin $\bf S$)
by the spin interaction $-\bf S \cdot B$.
The latter splitting implies the
existence of a supersymmetry of the
type discussed in refs.
[17].
This supersymmetry is not directly
relevant to the discussion here,
and the spin degree of freedom is neglected in what follows.

The combination of the cyclotron and axial motions forms
a system of two one-dimensional oscillators.
This becomes a physical realization
of a $D=2$ harmonic oscillator
when the applied electromagnetic fields are chosen
such that $\omega_c =\omega_z$,
i.e.,
$$ m V = \vert e \vert B^2 d^2
\quad .\eqno(51)$$
The quantum problem can then be separated in polar coordinates
and the radial equation has the general form of Eq.\ (19)
with
$\Gamma = -\half$,
and with suitable constant factors inserted
to allow for non-unit binding frequency and for SI units.

For fixed $L$,
this oscillator can be used
as the bosonic partner $H_+$ in a supersymmetric
quantum mechanics.
The partner hamiltonian $H_-$ is specified by Eq.\ (23).
It represents a system having an eigenspectrum degenerate
with the bosonic sector but with the ground state missing.
As in atomic supersymmetry,
one practical realization of this is to fill the ground
state with particles and invoke the Pauli principle.
If $L=0$, for example, $H_+$ describes the S orbitals of
the simplest geonium atom with one trapped particle.
Then,
$H_-$ can be interpreted as an effective theory
describing the behavior of the `valence' particle in
the S orbitals of a more complex geonium atom
in which the 1S (and 2P) orbitals are filled.
This interpretation invokes the approximation in which
particle interactions other than those implied
by the Pauli principle are disregarded.
All the atomic supersymmetries
of ref. [3] have analogues in this system.
For example,
there are connections between pairs of geonium atoms
throughout the geonium periodic table.

There exists a mapping
between the $d=3$ Coulomb problem
and the $D=2$ harmonic oscillator,
as discussed above.
The connection between the eigenfunctions given in Eq.\ (24)
therefore establishes a correspondence between
eigenfunctions of elements in the usual periodic table
and the geonium periodic table.
The map (24) is fixed here by setting $\lambda = 1$, so that
$$N=2n-1~~,~~~~L=2l+1
\quad .\eqno(52)$$
Moreover,
in the exact supersymmetry limit
this map induces other maps involving the supersymmetric partners
so that all four hamiltonians are linked.

The supersymmetries are broken by the interaction of the
valence particle with the `core' of the geonium atom.
These interactions will shift the eigenenergy of the
valence particle
from $E_N = N+1$
(in the level with quantum numbers $N,L$)
to some other energy
$$E_{N^\ast} = N^\ast +1
\quad ,\eqno(53)$$
where by definition
$$N^\ast = N - 2 \Delta (N,L)
\quad .\eqno(54)$$
A model incorporating these exact new eigenenergies
and yielding analytical solutions has been introduced
in the previous section.
In the present case,
it is obtained by adding an anharmonic term
to the oscillator hamiltonian,
giving the radial equation (43) with $\Gamma = -\half$
and with suitable dimension-correcting factors inserted.
The analytical solutions are
given by a corresponding modification of Eq.\ (45).

It is physically plausible
to conjecture that the quantum anharmonicity
rapidly approaches an asymptotic value as $N$ becomes large,
i.e.,
$$\Delta (N,L) \simeq \Delta (L)
\quad .\eqno(55)$$
In this case,
the eigensolutions form a complete and orthogonal set.
The previous section also provides
a mapping between this theory and
the analytical quantum-defect theory
for ordinary atoms.
Combined with the magnetron and spin splittings,
the anharmonic model is likely to provide a simple method
for calculations of physical properties of the
valence particle in geonium atoms.

\bigskip\bigskip
\leftline{\bf
ACKNOWLEDGMENTS
}
\bigskip

I thank
Robert Bluhm, Mike Nieto, Bob Pollock, Stuart Samuel,
and Rod Truax for discussion.
Part of this work was performed at the Aspen Center for Physics.
This research was supported in part by the United States
Department of Energy under contracts
DE-AC02-84ER40125 and DE-FG02-91ER40661.

\def\LMP #1 #2 #3 {Lett. Math. Phys. {\bf #1}, #3 (19#2).}
\def\CMP #1 #2 #3 {Comm. Math. Phys. {\bf #1}, #3 (19#2).}
\def\NCA #1 #2 #3 {Nouvo Cim. {\bf #1A}, #3 (19#2).}
\def\NCL #1 #2 #3 {Nuovo Cim. Lett. {\bf #1}, #3 (19#2).}
\def\NPA #1 #2 #3 {Nucl. Phys. {\bf A#1}, #3 (19#2).}
\def\NPB #1 #2 #3 {Nucl. Phys. {\bf B#1}, #3 (19#2).}
\def\PRA #1 #2 #3 {Phys. Rev. A {\bf #1}, #3 (19#2).}
\def\PRD #1 #2 #3 {Phys. Rev. D {\bf #1}, #3 (19#2).}
\def\PLB #1 #2 #3 {Phys. Lett. {\bf #1B}, #3 (19#2).}
\def\PRL #1 #2 #3 {Phys. Rev. Lett. {\bf #1}, #3 (19#2).}
\def\PTP #1 #2 #3 {Prog. Theor. Phys. {\bf #1}, #3 (19#2).}
\def\PL  #1 #2 #3 {Phys. Lett. {\bf #1}, #3 (19#2).}
\def\JMP #1 #2 #3 {J. Math. Phys. {\bf #1}, #3 (19#2).}
\def\PNA #1 #2 #3 {Proc. Natl. Acad. Sci. US {\bf #1}, #3 (19#2).}
\def\JPA #1 #2 #3 {J. Phys. A: Math. Gen. {\bf #1}, #3 (19#2).}
\def\PRR #1 #2 #3 {Phys. Rev. {\bf #1}, #3 (19#2).}

\bigskip\bigskip
\leftline{\bf
REFERENCES}
\bigskip

\item{1.}
F. Iachello, Phys. Rev. Lett. {\bf 44}, 772 (1980);
A. Balantekin, I. Bars and F. Iachello,
Nucl. Phys. {\bf A370}, 284 (1981).

\item{2.}
H. Nicolai, J. Phys. A {\bf 9}, 1497 (1976);
E. Witten, \NPB 188 81 513.

\item{3.}
V.A. Kosteleck\'y and M.M. Nieto,
Phys. Rev. Lett. {\bf 53}, 2285 (1984);
Phys. Rev. A {\bf 32}, 1293 (1985).

\item{4.}
V.~A.~Kosteleck\'y, M.~M.~Nieto, and D.~R.~Truax,
Phys.~Rev.~D {\bf 32}, 2627 (1985).

\item{5.}
J.~R.~Rydberg, Kongl. Sven. vetensk.-akad. hand. {\bf 23},
no. 11 (1890); Philos. Mag. {\bf 29}, 331 (1890).

\item{6.}
V.~A. Kosteleck\'y and M.~M. Nieto,
Phys. Rev. A {\bf 32}, 3243 (1985).

\item{7.}
J. Beckers and N. Debergh,
``On a Parastatistical Hydrogen Atom
and its Supersymmetric Properties,''
Li\`ege preprint, 1992.

\item{8.}
W.L. Wiese, M.W. Smith, and B.M. Glennon,
\it Atomic Transition Probabilities, Vols. 1 and 2, \rm
Natl. Bur. Stand. (U.S.)
Natl. Stand. Ref. Data Ser. Nos. 4 and 22
(U.S. GPO, Washington, D.C., 1966 and 1969).

\item{9.}
G.-W. Wen, ``Matrix Elements of the Radial Operators and their
Recursion Relations in Analytical Quantum Defect Theory,''
Hunan Normal preprint, 1992.

\item{10.}
M.T. Djerad, J. Phys. II {\bf 1}, 1 (1991).

\item{11.}
R.E.H. Clark and A.L. Merts,
J. Quant. Spectrosc. Radiat. Transfer {\bf 38}, 287 (1987).

\item{12.}
R. Bluhm and V.A. Kosteleck\'y,
Phys. Rev. A {\bf 47}, 794 (1993).

\item{13.}
M.L. Zimmerman, M.G. Littman, M.M. Kash, and D. Kleppner,
Phys. Rev. A {\bf 20}, 2251 (1979).

\item{14.}
V.~A. Kosteleck\'y, M.~M. Nieto,
and D.~R. Truax,~\PRA 38 88 4413

\item{15.}
F.M. Penning, Physica (Utrecht) {\bf 3}, 873 (1936);
H. Dehmelt, Rev. Mod. Phys. {\bf 62}, 525 (1990);
W. Paul, \it ibid., \rm 531;
N.F. Ramsey, \it ibid., \rm 541.

\item{16.}
For a review of the case of a single trapped particle,
see, for example, L.S. Brown and G. Gabrielse,
Rev. Mod. Phys. {58}, 233 (1986).

\item{17.}
R. Jackiw,
Phys. Rev. D {\bf 29}, 2375 (1984);
R. Hughes, V.A. Kosteleck\'y and M.M. Nieto,
Phys. Lett. B {171}, 226 (1986);
Phys. Rev. D {\bf 34}, 1100 (1986).
See also B.W. Fatyga, V.A. Kosteleck\'y,
M.M. Nieto and D.R. Truax,
Phys. Rev. D {\bf 43}, 1403 (1991).

\vfill\eject
\end